\documentclass[a4paper]{jpconf}
\usepackage{graphicx}
\begin{document}
\title{Numerical calculation of RPC time resolution}

\author{Jaydeep Datta $^{1,2,}$\footnote[3]{Present address:
Université libre de Bruxelles, Av. Franklin Roosevelt 50, 1050 Bruxelles}, Nayana Majumdar $^{1,2}$, Supratik Mukhopadhyay $^{1,2}$ and Sandip Sarkar $^{1,2}$}

\address{$^1$ Saha Institute of Nuclear Physics, Sector 1, AF Block, Bidhan Nagar, Bidhannagar, Kolkata, West Bengal 700064, India}
\address{$^2$ Homi Bhabha National Institute, 2nd floor, BARC Training School Complex, Anushaktinagar, Mumbai, Maharashtra 400094, India}

\ead{jaydeep.datta@ulb.be}

\begin{abstract}
Resistive Plate Chamber (RPC) is a gaseous detector, known for its good spatial resolution and excellent time resolution.
Due to its fast response and excellent time resolution, it is used for both triggering and timing purpose. 
But the time resolution of RPC is dependent on the detector geometry, applied voltage and the gas mixture used for detector operation. 
In this work, we have tried to develop a numerical model to estimate the time resolution of the detector.
The model is developed using COMSOL Multiphysics, a commercially available finite element method solver.
Using the primary ionization information from HEED and the electron transport properties from MAGBOLTZ, the model solves the Boltzmann equations to simulate the avalanche in the detector and finds the time to cross a previously determined threshold current, which is used to measure the time resolution of the detector.
\end{abstract}

\section{\label{sec:1}Introduction}
The era of new physics studies demands advanced detectors for precise measurements.
Different gaseous detectors like Drift Chamber, Resistive Plate Chamber (RPC), Gas Electron  Multiplier (GEM), Multi-gap RPC, Time Projection Chamber (TPC) have successfully played their roles in many contemporary experiments \cite{CMS:2006myw, ALICE:2000jwd, CBM MuCH:2000jwd, ICAL:2015stm}.
In many of these experiments, the timing information is very important to reconstruct both energy and direction of the particles produced in the elementary interactions in the detectors, which requires nanosecond or sub-nanosecond time resolution for this purpose.
RPC is a popular choice because it offers good spatial and temporal resolution and allows simplistic manufacture even in large dimension and rugged handling as additional advantages.
Its high efficiency and excellent time resolution have made it useful both as a trigger and timing device \cite{Abbrescia:1995uj}.
The sub nanosecond time resolution of this detector has drawn attention, and there have been many attempts to understand the working principle and characteristics of this device \cite{Moshaii:2012zz, Khorashad:2011zz, Bosnjakovic:2016zz, Bosnjakovic:2014zz, Riegler:2002vg, Lippmann:2003yb, Fonte2013, Datta:2020whg}.

RPC detects charged particles from the electron avalanche resulted from the multiplication of primary electrons and ions created by the incident particle through ionization of the molecules present in its gas medium.
The movement of the avalanche induces a current signal over the read-out electrodes placed outside the gas gap via capacitive coupling.
The standard deviation of the distribution of the signal arrival time is defined as time resolution of RPC which is determined by the avalanche formation followed by its development and propagation in the detector.
In this work, we have attempted to simulate the time resolution of RPC from hydrodynamic assumptions using a numerical model similar to the one described in \cite{Datta:2021llj}.
The model uses a commercially available finite element method solver, COMSOL Multiphysics \cite{comsol}, to solve the Boltzmann equations for charged particles in fluid to simulate avalanche development and propagation in RPC.
In doing so, this model uses the electron drift properties in gas medium calculated from MAGBOLTZ \cite{Biagi:1989rm} and primary ionization information from HEED \cite{Smirnov:2005yi}.
The hydrodynamic assumptions of the model averages out the inherent statistical fluctuation of the process.
So, in an attempt to incorporate the stochastic nature of this phenomenon, two sources of fluctuation, namely the number of primary electrons and their distribution in the gas gap, have been introduced in this model.
The work describes the procedure to numerically evaluate the time resolution of RPC at different applied high voltages using this simulation model.

The section \ref{sec:2} will describe the model, followed by the description of the parameters used in the same.
The calculation of time resolution will be discussed in section \ref{sec:3} and the simulation results in section \ref{sec:4}.
Further, the scopes of the work will be discussed in section \ref{sec:5}. 
\section{\label{sec:2}Numerical model}
The charged particles, passing through an RPC, ionize the molecules of the filling gas mixture of the detector.
The primary electrons and ions created in the ionization process accelerate under the applied electric field and undergo collisions with neutral gas molecules.
Some of these collisions are elastic and some are inelastic ones.
The inelastic collisions of the electrons with ions, and neutral gas molecules either create more electrons or help in recombination or absorption of them depending upon the existing electric field.
The loss of electrons per unit length in the gas mixture is known as the attachment coefficient, and the creation of them per unit length is called the Townsend coefficient.
When the Townsend coefficient is more than the attachment one, it leads to an avalanche of electrons.
The movement of the swarm of electrons under the applied electric field induces a current on the read-out strips, as mentioned earlier.
In an experiment, the time when the induced signal crosses a pre-defined threshold level, is recorded as signal arrival time.
The signal arrival time has a fluctuation following the statistical nature of the detector dynamics, which is denoted as the time resolution of the device.
The description of the numerical model developed to simulate the avalanche formation and its growth and propagation is available in \cite{Datta:2021llj}.
A brief discussion on the same has been furnished below.
\subsection{\label{sec2a}Model geometry}
Instead of a realistic 3D simulation, we have assumed a 2D geometry of RPC, as shown in the following figure \ref{fig:1}.
\begin{figure}[h]
\centering
\includegraphics[width=0.5\linewidth]{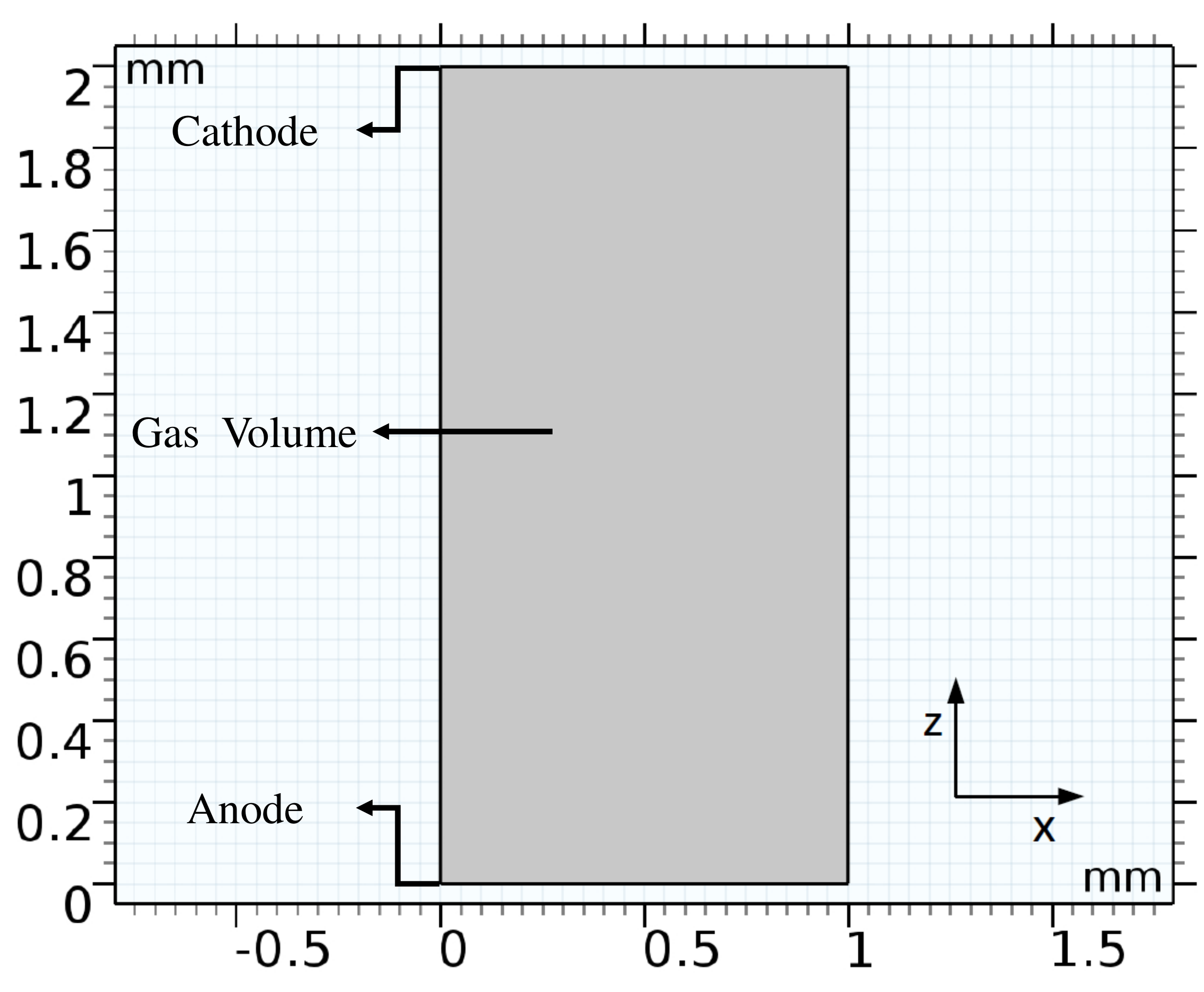}
\caption{Geometry used for simulation.\cite{Datta:2021odg}}
\label{fig:1}
\end{figure}
The geometry is 1 mm wide in the X-direction and 2 mm in Z-direction.
The 2 mm length in Z-direction represents the 2 mm gas gap of RPC.
A width of 1 mm along the X-direction has been chosen to minimize necessary computational resource without compromise on any important physical process.
The cathode has been placed to be at Z = 2 mm and the anode Z = 0 mm.
\subsection{\label{sec:2b}Simulation model}
It has been shown in \cite{Ammosov:1996gg} that the voltage applied to the resistive coating of RPC is realized by the gas gap without any significant loss.
So, in our model, the cathode and anode boundaries of the gas gap have been kept at the applied high voltages.
The model calculates the electric field in the geometry using the \emph{"Electrostatic"} module of the COMSOL \cite{comsol}.
While calculating the electric field, the module assumes a symmetry along the Y-direction.
As space charge effect has a significant impact on the growth of the RPC avalanche \cite{Riegler:2002vg, Lippmann:2003yb}, the same has been included in the temporal evolution of avalanche.
At each time step, the model calculates the electric field, $\vec E$, incorporating the instantaneous charge distribution including the space charge density, $\rho$, using the following equations \ref{eq9} and \ref{eq10}.
\begin{eqnarray}
\label{eq9}
\vec{E}=-\vec{\nabla} V\\
\label{eq10}
-\vec{\nabla}  d_y (\epsilon_0 \vec{\nabla} V - \vec{P}) = \rho\\
\label{eq11}
\rho = q_e(n_i -n_e).
\end{eqnarray}
Here V is the potential, q$_e$ is the magnitude of the charge of the electron, $d_y$ is the depth in the Y-direction, $\vec{P}$ is the polarization vector, $\epsilon_0$ is the permittivity of the vacuum.
Following the geometry, the calculated electric field is along the Z-direction.

The model uses hydrodynamic approximations to simulate the avalanche growth, assuming the medium as charged fluid.
This assumption leads to the following equations, to be solved to simulate the avalanche propagation in the RPC.
\begin{eqnarray}
	\label{eq1}
	\frac{\partial n_k}{\partial t} + \vec{\nabla} \cdot (-D_k \vec{\nabla} n_k + \vec {u}_k n_k) = R_k\\
	\label{eq2}
	R_k = S_e + S_{ph}\\
	\label{eq3}
	S_e = (\alpha (\vec{E}) - \eta (\vec{E})) |\vec{u}_e| n_e (\vec{x}, t)\\ 
	\label{eq4}
	S_{ph} = Q_e \mu_{abs} \psi_0.
\end{eqnarray}
Here $n_k$, ($k= i, e$), represents the concentration of the ions and electrons respectively while $D_k$, $\vec u_k$ and $R_k$ are their diffusion, drift velocity and rate of production, respectively.
As it is obvious from equation \ref{eq2}, $R_k$ is the sum of two source terms, $S_e$ and $S_{ph}$, namely the charges produced through Townsend ionization and photo-ionization mechanisms, respectively.
Here, $Q_e$ is the quantum efficiency of the filling gas for electron generation from photo-ionization, $\mu_{abs}$ is photo-absorption coefficient of the quencher component of the gas mixture, and $\psi_0$ is the photon flux generated in the detection volume.
The $\mu_{abs}$ has been calculated considering the corresponding  photo-absorption cross-section of the specific gas component obtained from relevant sources \cite{Lombos1967, Orlando1991}.
The electron drift velocity, diffusion coefficient, Townsend coefficient and attachment coefficient are calculated using MAGBOLTZ, which provides these parameters as a function of electric field only for electrons.
The drift velocity of ions has been set to three orders of magnitude smaller than that of electrons for all electric fields \cite{Raizer, deng}.
As ions are much heavier than the electrons, they will hardly diffuse in that small-time scale. 
So, the diffusion coefficient has been set to a very small value of 10$^{-9}$ m$^2$/sec.
Equations \ref{eq1} - \ref{eq4} are solved numerically using the \emph{'Transport of Diluted Species'} module of COMSOL Multiphysics.

The photo-ionization plays an important role in avalanche propagation,
The dynamics of the photons has been taken care of following
the equations available in \cite{Capeillere2008}.
\begin{eqnarray}
	\label{eq5}                   
	\vec{\nabla} (-c\vec{\nabla} \psi_0) + a \psi_0 = f\\   
	\label{eq6}
	c = \frac{1}{3 \mu_{abs}}\\
	\label{eq7}
	a = \mu_{abs}\\
	\label{eq8}
	f = \delta S_e
\end{eqnarray}
This set of equations are solved using \emph{"Coefficient Form Partial Differential Equation"} module of COMSOL \cite{comsol}.
Here, $\delta$ in equation \ref{eq8} represents the number of excited molecules for each ionized molecule.
\section{\label{sec:3}Parameters and calculation}
The hydrodynamic assumptions of this numerical model neglects the statistical fluctuations present in avalanche propagation, which is a stochastic process by nature.
Two important factors, which govern the avalanche evolution are, namely the number of total primary electron-ion pair and their spatial distribution in the gas gap.
In this simulation work, we have incorporated only the fluctuations for total number of primary electrons and their weighted mean Z-position.
However, the effects from electronics have not been included in this work.

\begin{figure}
\centering
\includegraphics[width=0.5\linewidth]{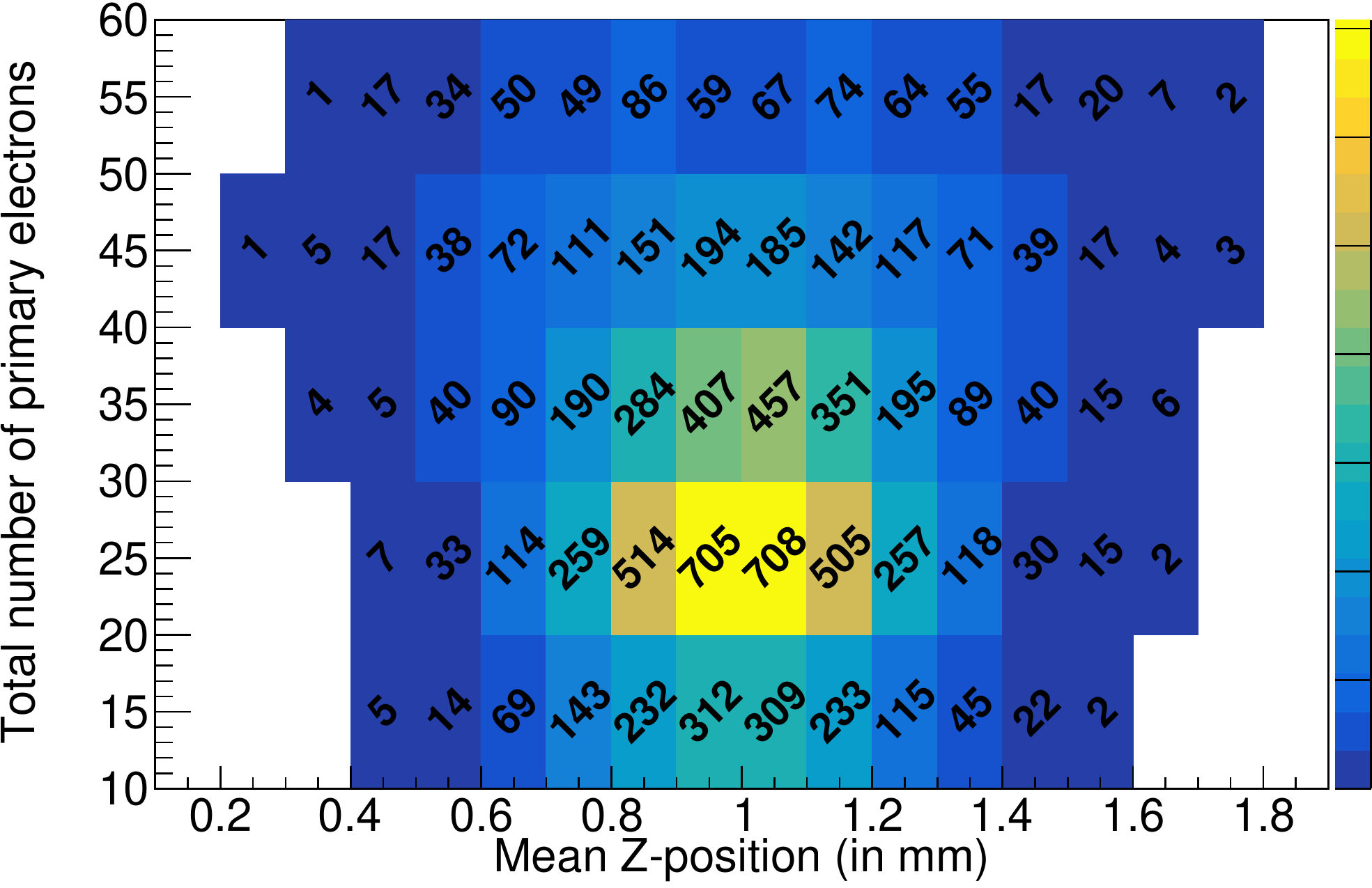}
\caption{Distribution of muon events based on total number of primary electrons and mean Z-position.\cite{Datta:2020whg}}
\label{fig:2}
\end{figure}

Using HEED simulation software, 10000 muon events following the atmospheric muon energy spectrum, have been simulated.
The muons were chosen to have a zenith angle between 0$^\circ$ and 13$^\circ$.
The gas medium of RPC has been chosen to be a mixture of R134a (95.2\%), iso-butane (4.3\%) and SF$_6$ (0.2\%) at 293.15 K and 1 bar pressure.
HEED provides the number of primary electrons and ions created in the gas and their position.
For each of the event, the total number of primary electrons and the position of each of the electrons have been stored.
According to the geometry, the applied electric field is along the positive direction of Z-axis and the muons have been considered to travel along the negative Z-axis.
For each event, from the primary ionization information, the weighted mean Z-position of the primary electrons has been calculated.
Using the two variables, total number of primary electrons and their mean Z-position, all the 10000 muon events have been distributed.
We have considered events with total number of primary electrons between 10 and 60 and mean Z-position between 0.1 mm and 1.9 mm, which covers 90\% of all the events.
Figure \ref{fig:2} shows the distribution of the events for the said gas mixture as function of the above-mentioned variables.
The bin size of the total number of electrons has been chosen to be 5 and that of the weighted mean Z-position has been chosen to be 0.1 mm.
So, for each of the combination of these two variables, simulations have been carried out.

The primary electron distribution has been assumed to be a two variable Gaussian distribution of Z-coordinate and X-coordinate.
The mean of Z-coordinate is the weighted mean Z-position, and the standard deviation for this is chosen in such a way that the five sigma of the distribution is bound by the 2 mm gas gap.
For the X-coordinate, the mean is 500 $\mu$m and the standard deviation is 0.01 mm.

The simulation model calculates the current due to the movement of the electrons in the gas gap using the Ramo's theorem \cite{Ramo:1939vr}.
In order to calculate the weighting field following the method described in \cite{Riegler:2002vg}, the electrode thickness has been chosen to be 2.8 mm and the relative permittivity of the electrode material has been considered as 8. 
In simulation, a current threshold of 1.25 $\mu$A has been chosen.
To find the signal arrival time when the induced current crosses the threshold, a stop condition has been implemented in the model, which stops the simulation at the time when the induced current crosses the threshold.
This time is stored for all the combinations to calculate the time resolution.
\section{\label{sec:4}Results}
\begin{figure}[h]
\centering
\begin{minipage}{0.49\linewidth}
\centering
\includegraphics[width=\linewidth]{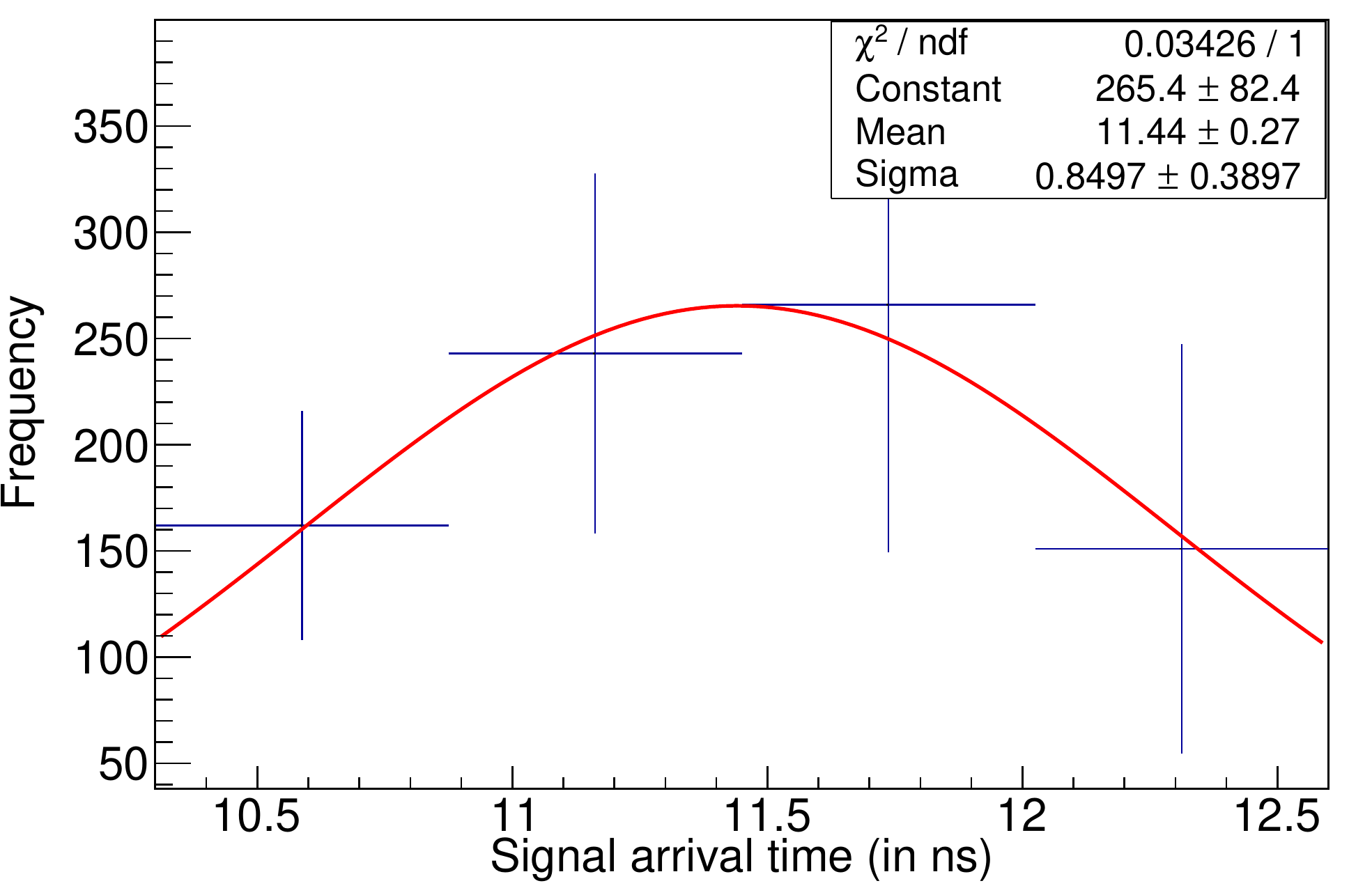}
\caption{Distribution of signal arrival time for electric field 44 kV/cm.}
\label{fig:3}
\end{minipage}%
\begin{minipage}{0.49\linewidth}
\centering
\includegraphics[width=\linewidth]{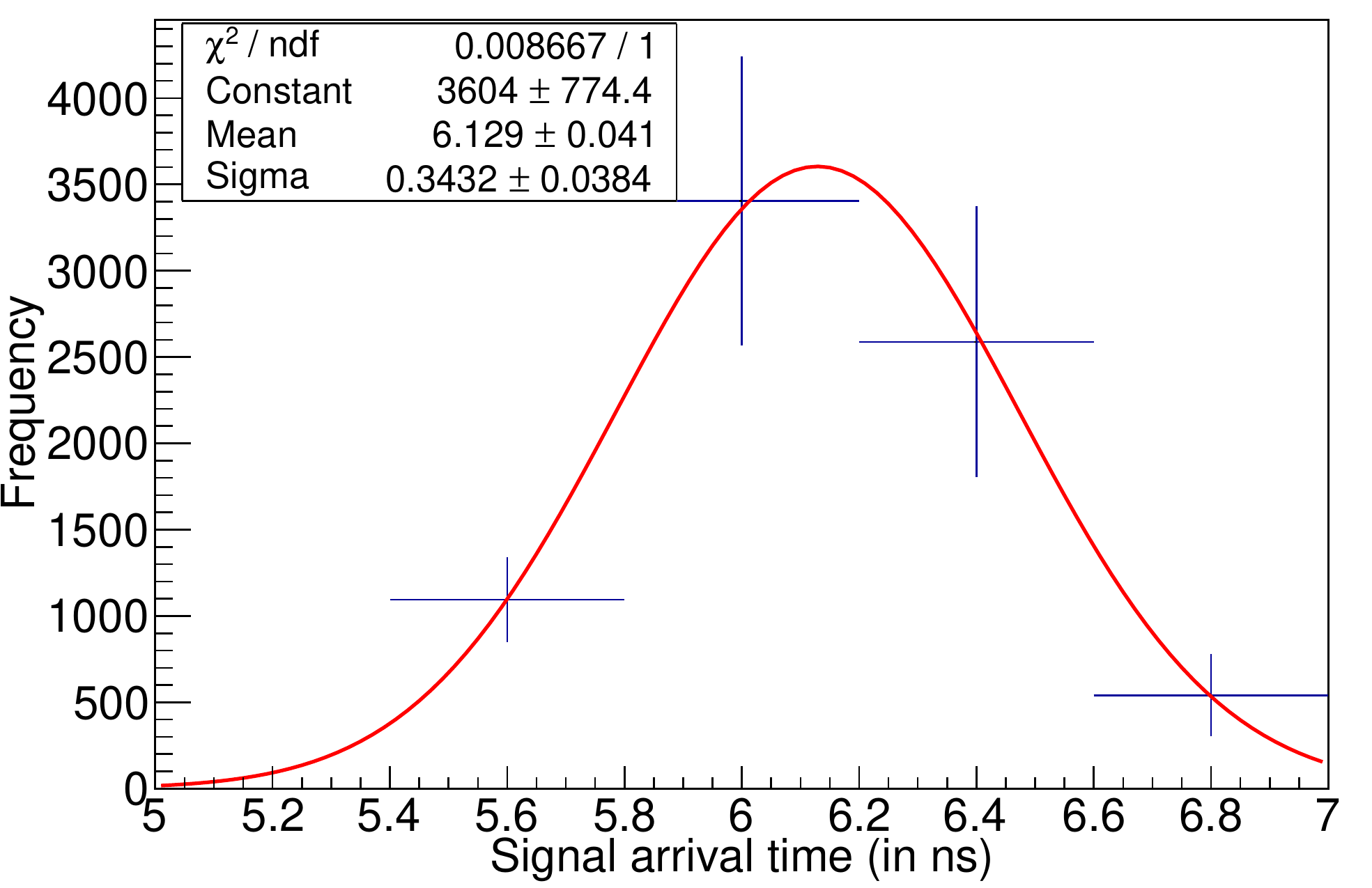}
\caption{Distribution of signal arrival time for electric field 49 kV/cm.}
\label{fig:4}
\end{minipage}
\end{figure}

From numerical simulation for each of the combination of number of primary electrons and the weighted mean Z-position, the time of threshold crossing is calculated.
This time is denoted as the signal arrival time.
Once we have got the signal arrival time for all the combinations, we have found out the occurrence frequency of this by adding up the number of events with the same signal arrival time. 
Using this information, the distribution of signal arrival time for each applied high voltage is plotted.
The distribution is fitted with a Gaussian distribution, whose standard deviation denotes the time resolution for that applied high voltage.
Figures \ref{fig:3} and \ref{fig:4} show the distributions, resulting fit and relevant parameters for two typical studies.

The time resolution for different high voltages is shown in table \ref{table1} where time resolution is quoted for different electric fields.
\begin{table}
	\centering
	\begin{tabular}{@{}ccc@{}}
		\hline
		Applied electric field (kV/cm) & Time resolution (this work) & Time resolution (\cite{Jash:2018cxf})\\
		\mr 
		 44 & 0.85 ns & 1.17 ns\\
		 46 & 0.63 ns & 0.49 ns\\
		 47 & 0.55 ns & 0.41 ns \\
		 48 & 0.43 ns & 0.36 ns\\
		 49 & 0.34 ns & 0.24 ns \\
		\br
	\end{tabular}
	\caption{\label{table1} Time resolution simulated in this work and by A. Jash et al. \cite{Jash:2018cxf} as a function of applied electric field.}
\end{table}
The numerically calculated time resolution has been compared with a previous numerical calculation by A. Jash et al. \cite{Jash:2018cxf}, where the authors numerically evaluated RPC time resolution for different applied electric fields as well as the different percentage of SF${_6}$ in the gas mixture.
Though there was no calculation with the gas mixture considered in the present work, we have compared our results to that obtained for the gas mixture containing 0.28\% of SF$_6$ which is close to our choice of our gas mixture.
The reason behind the disagreement between the results is likely to be due to the difference in the simulation method.
\section{\label{sec:5}Conclusion}
A simulation model based on hydrodynamics has been developed and used to calculate the time resolution of RPC.
For the gas mixture of R134a, iso-butane and SF$_6$, time resolution of an RPC with 2 mm gas gap has been calculated for different voltages and presented here.
At working voltage of an 2 mm gas gap RPC with this gas mixture, the calculated time resolution has been found around 340 ps.
To include some statistical nature of the physical processes, the fluctuation of total number of primary electrons and their distribution in the gas gap has been incorporated in a probabilistic way.
Inclusion of gain fluctuation and electronics effects in the presented model are planned in future to make the model more realistic.
\ack{}
The authors like to thank Saha Institute of Nuclear Physics and Homi Bhabha National Institute for their help.

\end{document}